\documentclass{article}

\PassOptionsToPackage{numbers, compress}{natbib}

\usepackage{tcolorbox}
\definecolor{cadmiumgreen}{rgb}{0.0, 0.42, 0.24}
\definecolor{oldmauve}{rgb}{0.4, 0.19, 0.28}
\definecolor{royalazure}{rgb}{0.0, 0.22, 0.66}
\definecolor{harvardcrimson}{rgb}{0.79, 0.0, 0.09}
\definecolor{lightmauve}{rgb}{0.86, 0.82, 1.0}
\definecolor{darkbrown}{rgb}{0.4, 0.26, 0.13}%
\usepackage[colorlinks = true,
            linkcolor = royalazure,
            urlcolor  = royalazure,
            citecolor = royalazure,
            anchorcolor = royalazure]{hyperref}

    \usepackage[preprint]{neurips_2024}

\usepackage[utf8]{inputenc} 
\usepackage[T1]{fontenc}    
\usepackage{hyperref}       
\usepackage{url}            
\usepackage{booktabs}       
\usepackage{amsfonts}       
\usepackage{nicefrac}       
\usepackage{microtype}      
\usepackage{xcolor}         
\usepackage{algorithm}
\usepackage{algorithmic}
\usepackage{float}
\usepackage{booktabs}
\usepackage{multirow}
\usepackage{capt-of}
\usepackage{wrapfig,lipsum,booktabs}

\usepackage{enumitem,graphicx}
\newtheorem{theorem}{Theorem}[section]

\usepackage[textsize=tiny]{todonotes}
\usepackage{enumitem} 
\usepackage{pifont}

\def\Dseen{\mathcal{D}_{seen}}

\def\Dsyn{\mathcal{D}_{syn}}
\def\Dbri{\mathcal{D}_{bright}}

\def\Dext{\mathcal{D}_{ext}}

\definecolor{darkgreen}{RGB}{0,120,0}
\definecolor{darkred}{RGB}{200,0,0}

\def\app{\textbf{Appendix}~}

\usepackage{amsmath,amsfonts,bm}

\def\eg{$e.g.$}

\def\1{\bm{1}}

\def\vtheta{{\bm{\theta}}}

\title{Revisiting the Auxiliary Data in Backdoor Purification}

\author{Shaokui Wei\textsuperscript{1} \quad Shanchao Yang\textsuperscript{1} \quad Jiayin Liu\textsuperscript{1} \quad Hongyuan Zha\textsuperscript{1,2}\\
\textsuperscript{1}The Chinese University of Hong Kong, Shenzhen, Guangdong, 518172, P.R. China\\
\textsuperscript{2}Shenzhen Key Laboratory of Crowd Intelligence Empowered Low-Carbon Energy Network
}

\def\app{\textbf{Appendix}}

\begin{document}

\maketitle
\begin{abstract}
Backdoor attacks occur when an attacker subtly manipulates machine learning models during the training phase, leading to unintended behaviors when specific triggers are present.  To mitigate such emerging threats, a prevalent strategy is to cleanse the victim models by various backdoor purification techniques. Despite notable achievements, current state-of-the-art (SOTA) backdoor purification techniques usually rely on the availability of a small clean dataset, often referred to as \textbf{auxiliary dataset}. However, acquiring an ideal auxiliary dataset poses significant challenges in real-world applications.
This study begins by assessing the SOTA backdoor purification techniques across different types of real-world auxiliary datasets. Our findings indicate that the purification effectiveness fluctuates significantly depending on the type of auxiliary dataset used. Specifically, a high-quality in-distribution auxiliary dataset is essential for effective purification, whereas datasets from varied or out-of-distribution sources significantly degrade the defensive performance. Based on this, we propose  Guided Input Calibration (GIC), which aims to improve purification efficacy by employing a learnable transformation. Guided by the victim model itself, GIC aligns the characteristics of the auxiliary dataset with those of the original training set. Comprehensive experiments demonstrate that GIC can substantially enhance purification performance across diverse types of auxiliary datasets. The code and data will be available via \href{https://github.com/shawkui/BackdoorBenchER}{https://github.com/shawkui/BackdoorBenchER}.
\end{abstract}    
\section{Introduction}
Backdoor attacks pose a concealed yet profound security risk to machine learning (ML) models, for which the adversaries can inject a stealth backdoor into the model during training, enabling them to illicitly control the model's output upon encountering predefined inputs. These attacks can even occur without the knowledge of developers or end-users, thereby undermining the trust in ML systems. As ML becomes more deeply embedded in critical sectors like finance, healthcare, and autonomous driving \citep{he2016deep, liu2020computing, tournier2019mrtrix3, adjabi2020past}, the potential damage from backdoor attacks grows, underscoring the emergency for developing robust defense mechanisms against backdoor attacks.

To address the threat of backdoor attacks, researchers have developed a variety of strategies \cite{liu2018fine,wu2021adversarial,wang2019neural,zeng2022adversarial,zhu2023neural,Zhu_2023_ICCV, wei2024shared,wei2024d3}, aimed at purifying backdoors within victim models. These methods are designed to integrate with current deployment workflows seamlessly and have demonstrated significant success in mitigating the effects of backdoor triggers \cite{wubackdoorbench, wu2023defenses, wu2024backdoorbench,dunnett2024countering}.  However, most state-of-the-art (SOTA) backdoor purification methods operate under the assumption that a small clean dataset, often referred to as \textbf{auxiliary dataset}, is available for purification. Such an assumption poses practical challenges, especially in scenarios where data is scarce. To tackle this challenge, efforts have been made to reduce the size of the required auxiliary dataset~\cite{chai2022oneshot,li2023reconstructive, Zhu_2023_ICCV} and even explore dataset-free purification techniques~\cite{zheng2022data,hong2023revisiting,lin2024fusing}. Although these approaches offer some improvements, recent evaluations \cite{dunnett2024countering, wu2024backdoorbench} continue to highlight the importance of sufficient auxiliary data for achieving robust defenses against backdoor attacks.

While significant progress has been made in reducing the size of auxiliary datasets, an equally critical yet underexplored question remains: \emph{how does the nature of the auxiliary dataset affect purification effectiveness?} In  real-world  applications, auxiliary datasets can vary widely, encompassing in-distribution data, synthetic data, or external data from different sources. Understanding how each type of auxiliary dataset influences the purification effectiveness is vital for selecting or constructing the most suitable auxiliary dataset and the corresponding technique. For instance, when multiple datasets are available, understanding how different datasets contribute to purification can guide defenders in selecting or crafting the most appropriate dataset. Conversely, when only limited auxiliary data is accessible, knowing which purification technique works best under those constraints is critical. Therefore, there is an urgent need for a thorough investigation into the impact of auxiliary datasets on purification effectiveness to guide defenders in  enhancing the security of ML systems. 

In this paper, we systematically investigate the critical role of auxiliary datasets in backdoor purification, aiming to bridge the gap between idealized and practical purification scenarios.  Specifically, we first construct a diverse set of auxiliary datasets to emulate real-world conditions, as summarized in Table~\ref{overall}. These datasets include in-distribution data, synthetic data, and external data from other sources. Through an evaluation of SOTA backdoor purification methods across these datasets, we uncover several critical insights: \textbf{1)} In-distribution datasets, particularly those carefully filtered from the original training data of the victim model, effectively preserve the model’s utility for its intended tasks but may fall short in eliminating backdoors. \textbf{2)} Incorporating OOD datasets can help the model forget backdoors but also bring the risk of forgetting critical learned knowledge, significantly degrading its overall performance. Building on these findings, we propose Guided Input Calibration (GIC), a novel technique that enhances backdoor purification by adaptively transforming auxiliary data to better align with the victim model’s learned representations. By leveraging the victim model itself to guide this transformation, GIC optimizes the purification process, striking a balance between preserving model utility and mitigating backdoor threats. Extensive experiments demonstrate that GIC significantly improves the effectiveness of backdoor purification across diverse auxiliary datasets, providing a practical and robust defense solution.

Our main contributions are threefold:
\textbf{1) Impact analysis of auxiliary datasets:} We take the \textbf{first step}  in systematically investigating how different types of auxiliary datasets influence backdoor purification effectiveness. Our findings provide novel insights and serve as a foundation for future research on optimizing dataset selection and construction for enhanced backdoor defense.
\textbf{2) Compilation and evaluation of diverse auxiliary datasets:}  We have compiled and rigorously evaluated a diverse set of auxiliary datasets using SOTA purification methods, making our datasets and code publicly available to facilitate and support future research on practical backdoor defense strategies.
\textbf{3) Introduction of GIC:} We introduce GIC, the \textbf{first} dedicated solution designed to align auxiliary datasets with the model’s learned representations, significantly enhancing backdoor mitigation across various dataset types. Our approach sets a new benchmark for practical and effective backdoor defense.

\section{Related work}
\textbf{Backdoor attacks.}
DNNs confront a significant security threat from backdoor attacks, which can undermine the reliability of these models. In such attacks, adversaries manipulate the model's output by embedding a specific trigger within the input data. The victim model behaves normally for benign inputs but produces attacker-controlled outcomes when it detects the predefined trigger. Backdoor attacks are typically categorized into two types based on the nature of their triggers: static-pattern and dynamic-pattern attacks. Early research, such as BadNets \citep{gu2019badnets} introduced fixed-pattern triggers—like a white square placed in a corner of an image—that could be easily identified and mitigated through visual inspection or algorithmic detection.  Consequently, subsequent studies have shifted focus towards more covert methods of injecting triggers. For instance, the Blended approach \citep{chen2017targeted}  integrates the trigger imperceptibly into the host image, enhancing its stealth. Dynamic-pattern backdoor attacks represent a more advanced form of this threat. Methods like  WaNet \citep{nguyen2021wanet}, LF \citep{zeng2021rethinking}, and SSBA \citep{li2021invisible} generate sample-specific triggers that blend seamlessly with the input, making them significantly harder to detect. Furthermore, some techniques aim to preserve the relationship between the semantic information of the input and the associated label, exemplified by works such as LC \citep{shafahi2018poison} and SIG \citep{barni2019new}, thereby increasing the subtlety and effectiveness of the attack.

\textbf{Backdoor defenses.}
As backdoor attacks garner growing attention from the research community, a comprehensive array of strategies has emerged to safeguard machine learning systems across their entire lifecycle \cite{wu2023defenses}. These defense mechanisms can be categorized into three primary phases: pre-training, in-training, and post-training.

This paper centers on post-training defenses aimed at eliminating backdoors within pre-trained models. To this end, a prominent strategy involves the identification and removal of neurons linked to backdoor activities. Techniques such as FP \cite{liu2018fine}, ANP \cite{wu2021adversarial} and CLP \cite{zheng2022data}. An alternative method is to fine-tune the affected model to diminish backdoor influence. Key strategies here include NC \cite{wang2019neural}, i-BAU \cite{zeng2022adversarial}, NPD \cite{zhu2023neural}, and SAU \cite{wei2024shared}, which leverage adversarial techniques to simulate potential backdoor triggers, followed by fine-tuning to enhance the model's resistance against these reconstructed poisoned samples, thereby cleansing it from backdoor effects. In addition, NAD \cite{li2021neural}  utilizes a teacher network to guide the fine-tuning of a compromised student network, aiding in the mitigation of backdoors. The FT-SAM approach \cite{Zhu_2023_ICCV} and FST \cite{min2024towards} further advances fine-tuning  for backdoor mitigation.

Despite their notable defensive success, most SOTA purification techniques operate under the assumption that an auxiliary dataset is accessible, which can be a significant limitation in practical applications. To address this challenge, some researchers have turned their attention to filtering a clean dataset from poisoned dataset~\cite{zeng2022sift}, reducing the size of required auxiliary dataset~\cite{chai2022oneshot,li2023reconstructive, Zhu_2023_ICCV}, or even developing data-free methods~\cite{zheng2022data,hong2023revisiting}. While effective against certain attacks, these methods often encounter limitations when dealing with diverse models or more sophisticated attacks~\cite{dunnett2024countering,wu2024backdoorbench}. In summary, robust backdoor purification continues to rely heavily on the availability of adequate auxiliary data, underscoring the urgent need to develop practical and adaptive purification techniques.

\section{Unveiling the Power and Pitfalls of Auxiliary Dataset}

\subsection{Problem Setup}
\label{sec::pre}
\textbf{Notations.} 
In this paper, we focus on the image classification task, where each sample $\mathbf{x} \in \mathcal{X}$ is associated with a label $y \in \mathcal{Y}$. A deep neural network (DNN) model $f_{\vtheta}$, parameterized by $\vtheta$, is employed to classify $\mathbf{x}$. The label set $\mathcal{Y} = \{1, \dots, K\}$ represents the space of possible classes ($K \geq 2$), and $\mathcal{X}$ denotes the input sample space.

\textbf{Threat model.}
We consider a scenario in which an adversary manipulates a subset of the training data, embedding a hidden trigger to compromise the model. As a result, the victim model behaves normally on clean inputs but misclassifies any input containing the trigger $\Delta$ into a predefined target class $\hat{y}$. The proportion of manipulated samples in the training dataset is referred to as the \textbf{poisoning ratio}.

\textbf{Defender's goal and capabilities.}
The defender’s objective is to purify the victim model, mitigating the backdoor effect while preserving classification performance on clean, non-compromised inputs. This requires achieving an optimal balance between maintaining the model’s utility for its intended task and ensuring robustness against backdoor attacks. The defender has no prior knowledge of the backdoor trigger $\Delta$ or the target label $\hat{y}$, making the purification process more challenging. Additionally, the defender is provided with an auxiliary dataset, which is small and insufficient for training a new model from scratch.

\begin{table}[ht]
\centering
\caption{Classification of auxiliary datasets. The datasets can be categorized into two main types: those that the model has \textbf{seen} during its training phase and those that remain \textbf{unseen}. For datasets that are \textbf{unseen}, we further divide them into two subcategories:
\textbf{1) In-distribution data}, which typically includes meticulously chosen samples split from training or test datasets, and
\textbf{2) Out-of-distribution data}, encompassing synthetic datasets or external datasets sourced from outside the original dataset (\eg, web-based collections).
Some relevant papers utilizing these auxiliary datasets are also listed for reference.
}
\label{overall}
\setlength{\tabcolsep}{4pt} 
\renewcommand{\arraystretch}{1.2} 
\scalebox{0.75}{

\begin{tabular}{llll}
\toprule
Category                & Type                                 & Source                                                   & Work \\ \midrule
\multirow{3}{*}{Seen}   & \multirow{3}{*}{In-distribution}     & \multirow{3}{*}{Filtered from training dataset}    & \citet{wang2019neural, wu2021adversarial,Zhu_2023_ICCV}     \\ 
                        &                                      &                                                          & \citet{li2021neural, zhu2023neural,dunnett2024countering}\\ 
                        &                                      &                                                          & \citet{wei2024shared,wei2024d3, wubackdoorbench, wu2024backdoorbench}...\\ \midrule
\multirow{4}{*}{Unseen} & In-distribution                      & Filtered \& split from training/test datasets...       & \citet{wang2019neural,zeng2022adversarial,liu2018fine}...     \\ \cmidrule(l){2-4}
                        & \multirow{3}{*}{Out-of-distribution} & Synthetic datasets                                       &  \citet{wei2024shared,wei2024d3}...    \\\cmidrule(l){3-4}
                        &                                      & External datasets            &       \\ 
                        \bottomrule 
\end{tabular}}
\end{table}

\textbf{Classification of auxiliary datasets.}
As summarized in Table~\ref{overall}, we classify auxiliary datasets into two main categories based on their relationship to the training process of the victim models:
\begin{itemize}[leftmargin=*, topsep=2pt, itemsep=2pt, partopsep=2pt, parsep=2pt]
    \item \textbf{Seen data:} This category encompasses datasets that the victim models have been exposed to during the training phase. Such auxiliary datasets can be constructed by human inspection or applying some data filtering techniques \cite{chen2019detecting,zeng2022sift, zhu2023vdc} to the training datasets.
    \item \textbf{Unseen data:} Unseen dataset consist of data points that the model has not encountered during training. We further divided them into  two types:
    
    \ding{169} In-distribution data: These datasets contain samples that mirror the distribution of the training data but were excluded from the training process. Datasets in this type include but not limited to carefully selected subsets split from the original training or testing datasets.

    \ding{169} Out-of-distribution data: This type includes synthetic datasets\footnote{Advanced generative models may produce synthetic data that closely resembles in-distribution samples, but we still categorize these as OOD samples for simplicity and consistency.}, as well as external datasets sourced from external environments such as collecting from website. Out-of-distribution (OOD) data may significantly differ in characteristics from the training data, providing a challenge for backdoor purification.

\end{itemize}

\subsection{Role of Auxiliary Data in Backdoor Purification}
\label{sec::impact}

To analyze the role of auxiliary datasets in backdoor purification, we describe our experimental setup, including the dataset configurations, backdoor attack types, purification techniques, and evaluation metrics. All experiments are conducted on the standardized platform BackdoorBench \citep{wubackdoorbench, wu2024backdoorbench} to ensure fair comparisons.

\textbf{Datasets.} We conduct our experiments on three widely-used benchmark datasets: CIFAR-10 \citep{krizhevsky2009learning}, GTSRB \citep{stallkamp2011german}, and Tiny ImageNet \citep{le2015tiny}. 
For each dataset, we partition the original training set into two subsets:
\ding{182} \textbf{Training dataset} $\mathcal{D}_{\text{tr}}$, consisting of 95\% of the data, used for crafting poisoned dataset and training victim models. \ding{183} \textbf{Reserved dataset} $\mathcal{D}_{\text{unseen}}$, comprising the remaining 5\% in-distribution data and excluded from training.
To emulate practical defense scenarios, we construct clean \ding{184} \textbf{Seen dataset} $\Dseen$, derived from the poisoned dataset. For simplicity and clarity, $\Dseen$ is treated as an oracle dataset, assuming perfect knowledge of poisoning labels. This set allows us to assess the upper bound of backdoor purification performance when such dataset is available. Additionally, we generate \ding{185} \textbf{Synthetic dataset} $\Dsyn$, using generative models \citep{ho2020denoising,brock2018large}, enabling us to assess the utility of synthetic data in backdoor defenses.
To investigate the impact of more out-of-distribution data, we introduce \ding{186}\textbf{Brightness-transformed dataset} $\Dbri$, generated by applying brightness transformations which were intentionally excluded during training.This mimics auxiliary datasets collected under varying lighting conditions. For CIFAR-10, we include an \ding{187} \textbf{External dataset }$\Dext$, constructed by downsampling and selecting samples from ImageNet \citep{russakovsky2015imagenet} following the methodology in \citep{darlow2018cinic}, ensuring a diverse and realistic evaluation setting. More details of the datasets and visualization are provided in \app.

\textbf{Attack settings.} 
We evaluate seven widely studied backdoor attacks, including six dirty-label attacks (BadNets~\citep{gu2019badnets}, Blended~\citep{chen2017targeted}, WaNet~\citep{nguyen2021wanet}, Low Frequency  (LF)~\cite{zeng2021rethinking}, Input-aware~\citep{nguyen2020input} and Sample-Specific Backdoor Attack (SSBA)~\citep{li2021invisible}), and one clean-label attack (Sinusoidal Signal (SIG)~\citep{barni2019new}). By default, all attacks were executed with a 10\% poisoning rate, targeting the $0^{th}$ class unless otherwise specified. To provide a thorough analysis, we use two popular neural network architectures: PreAct-ResNet18 \citep{he2016identity} and VGG19-BN \citep{simonyan2014very}.

\textbf{Backdoor purification.}
We consider seven recent advanced backdoor defense methods, including two model modification techniques ANP \cite{wu2021adversarial} and NPD \cite{zhu2023neural},  five tuning-based methods (vanilla fine-tuning (FT), FT-SAM \cite{Zhu_2023_ICCV}, FST \cite{min2024towards}, NAD \citep{li2021neural}) and SAU \citep{wei2024shared}). To isolate the impact of auxiliary datasets, we standardize the size of the auxiliary dataset to 5\% of the original training dataset. This setting, in line with previous studies, is sufficient for most purification methods to demonstrate their effectiveness. \textbf{Remark}: Due to space constraints, we present only the most representative results here. Additional experimental findings, including results for VGG19-BN and other datasets such as Tiny ImageNet, are provided in the \app.

\textbf{Metrics.}
The effectiveness of each defense method is assessed using two key metrics: Accuracy on Clean Data (\(\textbf{ACC}\)) and Attack Success Rate (\(\textbf{ASR}\)). \(\textbf{ACC}\) evaluates the model's performance on unaltered samples, \(\textbf{ASR}\) quantifies the proportion of poisoned samples misclassified to the target label chosen by the attacker. An effective defense should maintain high \(\textbf{ACC}\), while minimizing \(\textbf{ASR}\) values, ensuring both model utility and robustness.

\begin{figure*}[h]
    \centering
    \includegraphics[width=1\linewidth]{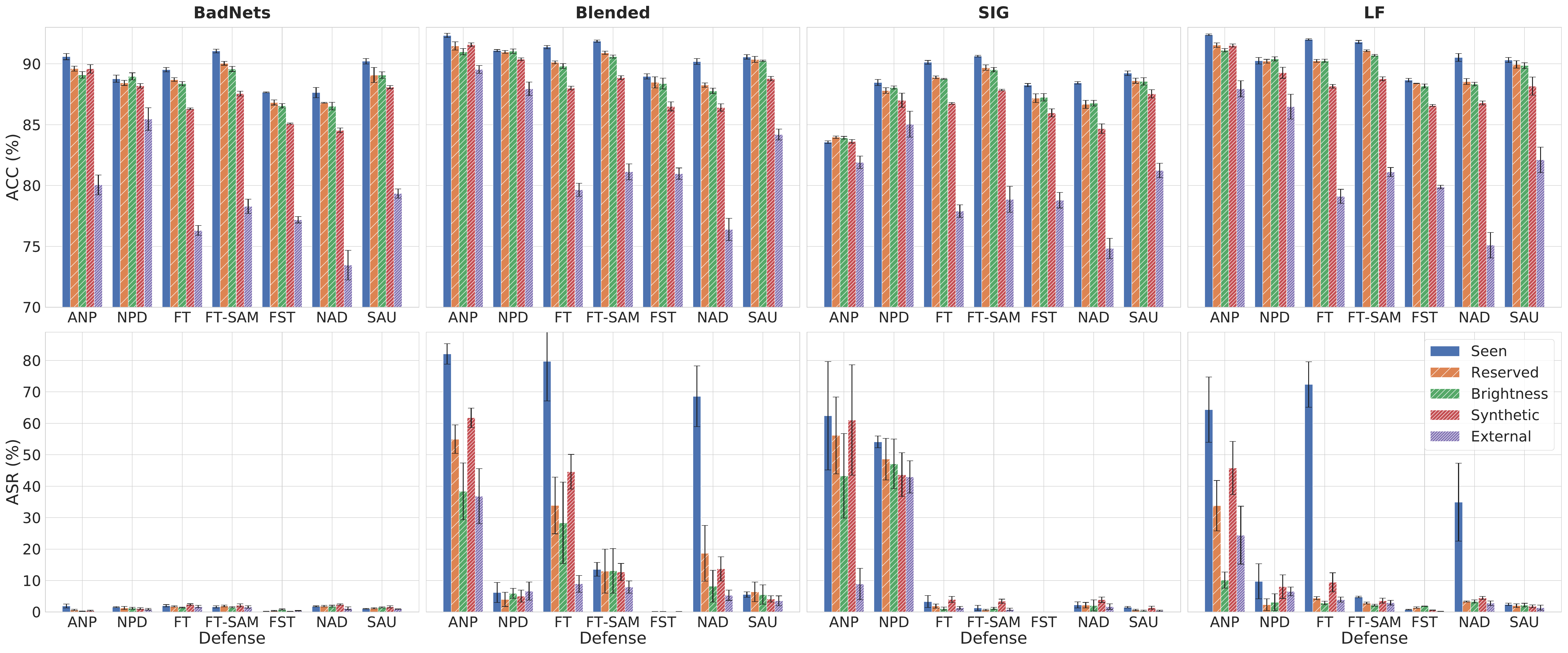}
    \vspace{-0.3in}
    \caption{Performance of backdoor purification techniques equipped with different types of auxiliary dataset. Each experiment is run five times and the average value with error bar is reported. Results on more attacks and purification techniques are provided in \app.}
    \label{fig:curves}
    \vspace{-0.2in}
\end{figure*}

We conducted each experiment five times and report the average performance with error bars in Figure~\ref{fig:curves}. The results provide critical insights into the role of auxiliary datasets in backdoor purification:

\ding{182} \textbf{Semantic alignment drives clean accuracy preservation:} Auxiliary datasets derived from the training dataset (i.e., $\Dseen$) consistently achieve the highest clean accuracy across all defense techniques. This highlights the pivotal role of semantic alignment with the model’s training distribution in preserving performance on clean samples. In contrast, auxiliary datasets that introduce semantic shifts, such as brightness-transformed dataset ($\Dbri$), synthetic dataset ($\Dsyn$), or external dataset ($\Dext$), bring significant performance degradation in ACC. This suggests that models perform optimally when they receive auxiliary data that aligns with the  data they were trained on.

\ding{183} \textbf{Unseen data can facilitate purification:} This advantage is especially evident in scenarios where seen dataset struggles, such as fine-tuning (FT) defenses against Blended attacks. The likely explanation is that the semantic shift introduced by unseen data pushes models away from the backdoored initializations, effectively reducing the attack’s effectiveness. These findings align with recent studies \citep{Zhu_2023_ICCV, wei2024backdoor}, which suggest that significant deviations from the model’s initial weights, such as those introduced by unseen data, can aid in backdoor purification can aid backdoor purification.  
\emph{However, the effect of unseen data is not universally beneficial.} For attacks that are already mitigated well by seen data, the introduction of unseen dataset can sometimes increase the variability in ASR,  leading to either further reductions or unexpected increases. This variability underscores the complex relationship between the diversity of auxiliary data and the effectiveness of purification methods, highlighting that unseen data may not always yield consistent improvements.

\ding{184} \textbf{Purification techniques exhibit different sensitivity to auxiliary dataset variations}: The impact of auxiliary dataset characteristics varies significantly across backdoor defense techniques. Specifically, model modification techniques, such as ANP  \cite{wu2021adversarial} and NPD  \cite{zhu2023neural}, which incur limited changes to model parameters exhibit slightly less sensitivity to auxiliary dataset, showing less reduction in ACC when unseen auxiliary dataset is employed. In contrast, fine-tuning-based methods, which involve more substantial adjustments to the model's weights, are highly sensitive to the quality of the auxiliary data.  While they effectively reduce ASR,  they often experience notable ACC degradation when the auxiliary dataset diverges significantly from the training distribution. This sensitivity underscores the importance of carefully selecting auxiliary datasets when using fine-tuning-based methods for backdoor purification.

\textbf{Investigation in latent space.} To gain deeper insights into the role of auxiliary datasets in backdoor purification, We analyze the latent space representations of the victim model. Specifically, we visualize the feature distributions of both the seen dataset and various auxiliary datasets using t-SNE, which allows us to highlight the shifts introduced by the auxiliary data.
\begin{figure}[h]
    \centering
    \includegraphics[width=0.5\linewidth]{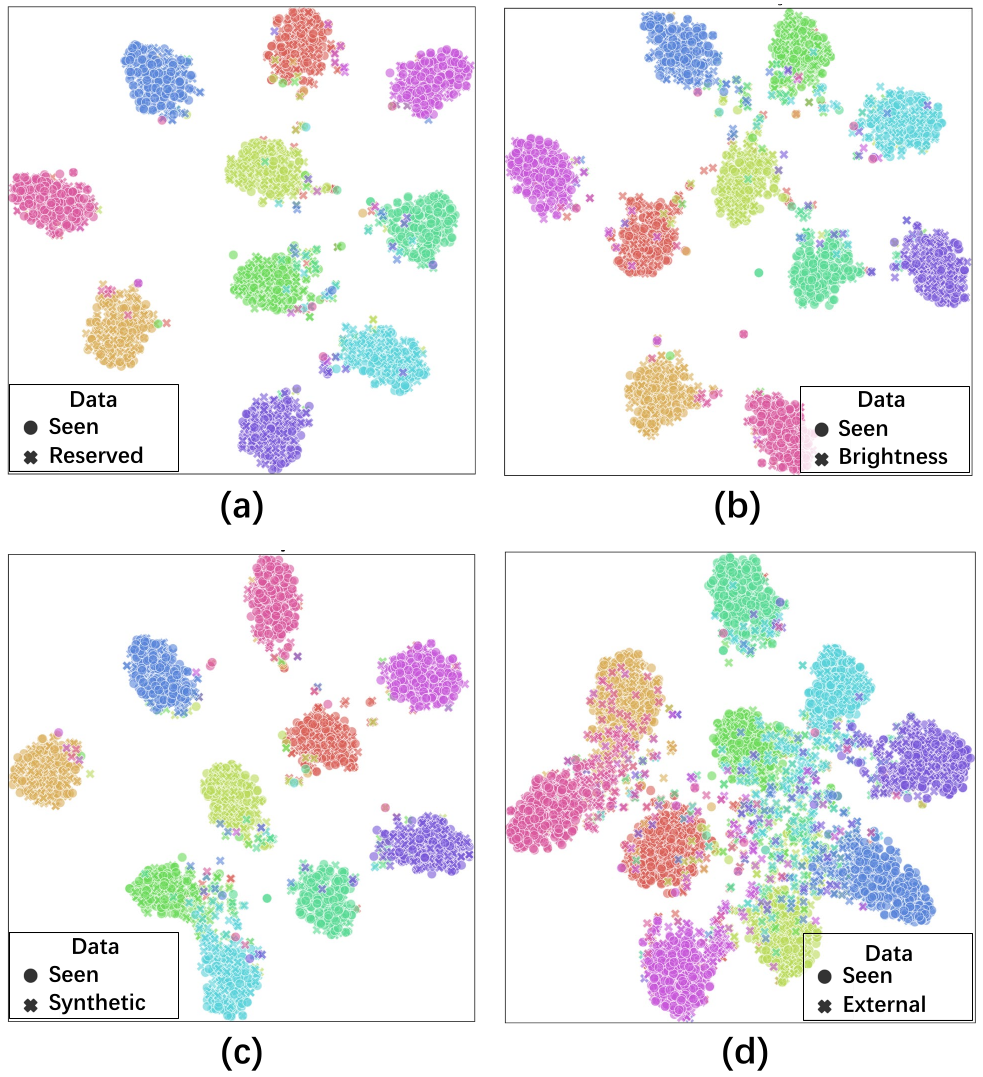}
    \caption{t-SNE visualization of features from different auxiliary datasets. In each plot, points representing different classes are depicted in distinct colors. Features from the seen dataset are marked with dots, while those from other datasets are represented with crosses.}
    \label{tsnep}
\end{figure}
We conduct experiments using PreAct-ResNet-18 on a victim model compromised by the BadNets attack. As summarized in Figure~\ref{tsnep}, we compare the feature distributions of the seen dataset with those of each unseen dataset. The t-SNE visualizations reveal that while the victim model can effectively extract features from diverse datasets, unseen auxiliary data introduce noticeable shifts in the feature space. This effect is particularly pronounced when the auxiliary dataset diverges from the original training distribution. For instance, brightness-transformed data exhibits a moderate deviation, indicating that even simple transformations impact the model’s  feature representations. Synthetic and external datasets show the most significant shifts, indicating that semantic transformations substantially alter the feature.

\textbf{In summary,} These findings emphasize the sensitivity of backdoor purification techniques to the choice of auxiliary datasets. Datasets with substantial semantic shifts can degrade model accuracy and destabilize attack success rates, underscoring the importance of carefully selecting high-quality auxiliary datasets.

\section{Enhancing Backdoor Purification with Guided Input Calibration}
\subsection{Guided Input Calibration}
\textbf{Motivation.} From our earlier analysis, we identified that auxiliary datasets, especially those that introduce significant semantic shifts, can degrade the model’s clean accuracy and destabilize attack success rates. This is particularly problematic in backdoor defense, where the goal is to both \textbf{remove the backdoor} and \textbf{maintain high clean accuracy}. Therefore, it is crucial that auxiliary datasets used for purification do not introduce large deviations from the model's training data, especially when substantial updates to model parameters are involved.

\textbf{Methodology.} The core of Guided Input Calibration is a learnable transformation function $g$, which adjusts the auxiliary dataset $\mathcal{D}_{aux}$ to better align with the model's expectations. The transformation function $g$ is designed to ensure that the adjusted data closely resembles the original training data distribution. This transformation is flexible and can take various forms, such as learnable perturbations \cite{song2023deep}, visual prompts \cite{bahng2022exploring}, or spatial transformations \cite{xiao2018spatially}.

To ensure broad applicability, we consider a scenario where no clean samples from the original training dataset are available for reference. This scenario is particularly challenging due to the lack of clean features, and it reflects practical settings. Instead of directly using clean data, the optimization of $g$ is guided by the \textbf{victim model}. Since the victim model is trained to predict clean samples correctly, the model’s output provides guidance for the calibration process. Specifically, the optimization objective for $g$ is defined as follows:
\begin{equation}
\min_{g:\|g(x_i)-x_i\|_p\leq \delta}  \sum_{(x_i,y_i)\in\mathcal{D}_{aux}} \ell(f(g(x_i)), y_i),
\end{equation}
where $\ell$ is the classification loss function (e.g., cross-entropy loss), and $\delta$ is a hyperparameter controlling the distance between the original sample $x_i$ and its calibrated version $g(x_i)$. The introduction of $\delta$ serves as a constraint on the transformation process, ensuring that $g(x)$ remains close to the original sample and does not transform the input into arbitrary patterns, especially those resembling trigger patterns with significant modifications. We empirically demonstrate in the \app that the transformed images do not exhibit characteristics of poisoned data, even for attacks with minor image modifications.

Through this objective, the transformation function $g$ is optimized to minimize classification error, effectively aligning the auxiliary data with the victim model's expectations.



\paragraph{Theoretical analysis.}
Here, we provide a theoretical foundation for the GIC. To simplify the analysis, we focus on a binary $(0-1)$ classification model $f$, which consists of a feature extractor $\phi$ and a linear classifier $W$. Given the training dataset $\mathcal{D}_{tr}$, the model $f$ is trained to minimize the Binary Cross-Entropy (BCE) loss:  
\begin{equation}
    \label{bce1}
    \min_{f} -\frac{1}{|\mathcal{D}_{tr}|} \sum_{(x_i, y_i) \in \mathcal{D}_{tr}} \left( y_i \log(p_i) + (1 - y_i) \log(1 - p_i) \right),
\end{equation}  
where $p_i = P(f(x_i) = 1) = \sigma(W^\top \phi(x_i))$ and $\sigma$ is the sigmoid activation function.  

We assume that the feature vectors have finite norms and let \( M =\max_{x} \|\phi(x)\|\) denote the largest norm across all feasible inputs. This assumption is practical and commonly encountered in machine learning models, especially neural networks. In practice, techniques such as input normalization, regularization (e.g., weight decay), and normalization layers (e.g., batch normalization) ensure that the features produced by the model are bounded, preventing arbitrary growth during training and inference.

In the GIC framework, the transformation $g(x)$ is trained to increase the prediction confidence of $f(g(x))$ for its correct label. This optimization naturally encourages $g(x)$ to resemble one of the samples in $\mathcal{D}_{tr}$, as training samples are the most confident predictions of $f$ after training. Let $x' \in \mathcal{D}_{tr}$ be such a sample whose confidence matches that of $g(x)$, i.e., $P(f(g(x)) = 1) = P(f(x') = 1)$.  

We now present the following theorem:  

\begin{theorem}
\label{thm}
Consider a model $f$ trained by solving the optimization problem in Equation~\ref{bce1}.  Under the assumption of bounded feature norm, the distance between the feature vectors $\phi(g(x))$ and $\phi(x')$ is bounded as:  
\[
\|\phi(g(x)) - \phi(x')\|^2 \leq 4M^2 - \frac{4}{\|W\|^2} \left( \log\left( \frac{1 - p}{p} \right) \right)^2,
\]  
where $p = P(f(g(x)) = 1)= P(f(x') = 1)$.
\end{theorem}

This theorem shows that as $g(x)$ is optimized to increase the prediction confidence $p$ for its correct label, the transformation encourages $g(x)$ to align with training sample $x'$ that has a similar high-confidence prediction. Specifically, the distance between the feature representations $\phi(g(x))$ and $\phi(x')$ is bounded by the equation above, which ensures that the transformation $g(x)$  does not drastically deviate from the training dataset. This alignment ensures that the auxiliary dataset, transformed by $g(x)$, produces features that are similar to those in the original training dataset. By promoting such alignment, GIC effectively ensures that the transformed auxiliary dataset remains consistent with the original training distribution, benefiting backdoor purification techniques. 

\subsection{Experiments}
\begin{table*}[ht]
\centering
    \caption{Experiments evaluating backdoor purification techniques with various auxiliary datasets on the BadNets attack. The table reports the accuracy (ACC) and attack success rate (ASR) for auxiliary datasets with and without applying Guided Input Calibration. Performance improvements are highlighted in darkgreen, while degradations are marked in darkred.}
    \label{gic_badnet}
    \setlength{\tabcolsep}{4.00pt} 
    \renewcommand{\arraystretch}{1} 
    \scalebox{0.7}{
    \begin{tabular}{c|cc|cc|cc|cc|cc|cc}
    \toprule
          Defense   $\rightarrow$  & \multicolumn{4}{c|}{ANP} & \multicolumn{4}{c|}{NPD}  & \multicolumn{4}{c}{FT} \\ \midrule
          GIC $\rightarrow$    & \multicolumn{2}{c|}{w/o GIC}   & \multicolumn{2}{c|}{w/ GIC}                                                                            & \multicolumn{2}{c|}{w/o GIC}                       & \multicolumn{2}{c|}{w/ GIC}                                                                            & \multicolumn{2}{c|}{w/o GIC}                       & \multicolumn{2}{c}{w/ GIC}                                                                            \\ \midrule
             Datasets $\downarrow$  & {ACC} & {ASR} & ACC                                               & ASR                                               & {ACC} & {ASR} & ACC                                               & ASR                                               & {ACC} & {ASR} & ACC                                               & ASR                                               \\ \midrule
    Reserved   & 89.03                   & 1.28                    & 89.71(\textcolor{darkgreen}{0.68}) & 0.10(\textcolor{darkgreen}{-1.18}) & 87.85                   & 0.20                     & 88.98(\textcolor{darkgreen}{1.13}) & 0.34(\textcolor{darkred}{0.14})    & 88.72                   & 1.90                     & 89.05(\textcolor{darkgreen}{0.33}) & 1.50(\textcolor{darkgreen}{-0.40}) \\
    Brightness & 86.91                   & 0.28                    & 88.93(\textcolor{darkgreen}{2.02}) & 0.06(\textcolor{darkgreen}{-0.22}) & 87.58                   & 0.38                    & 89.45(\textcolor{darkgreen}{1.87}) & 0.44(\textcolor{darkred}{0.07})    & 88.23                   & 1.57                    & 88.56(\textcolor{darkgreen}{0.33}) & 1.47(\textcolor{darkgreen}{-0.10}) \\
    Synthetic  & 88.44                   & 0.46                    & 89.48(\textcolor{darkgreen}{1.04}) & 0.09(\textcolor{darkgreen}{-0.37}) & 87.30                    & 1.00                       & 88.73(\textcolor{darkgreen}{1.43}) & 0.74(\textcolor{darkgreen}{-0.26}) & 86.19                   & 3.23                    & 87.60(\textcolor{darkgreen}{1.41}) & 1.72(\textcolor{darkgreen}{-1.51}) \\ 
    External   & 82.74                   & {0.00}   & 86.48(\textcolor{darkgreen}{3.74}) & 0.01(\textcolor{darkred}{0.01})    & 84.33                   & 0.11                    & 86.06(\textcolor{darkgreen}{1.73}) & 0.08(\textcolor{darkgreen}{-0.03}) & 76.42                   & 1.67                    & 85.06(\textcolor{darkgreen}{8.64}) & 1.14(\textcolor{darkgreen}{-0.52}) \\ \midrule \midrule
          Defense $\rightarrow$     & \multicolumn{4}{c|}{FT-SAM}                                                                                                                                & \multicolumn{4}{c|}{FST}                                                                                                                                   & \multicolumn{4}{c}{SAU}                                                                                                                                   \\ \midrule
        GIC $\rightarrow$       & \multicolumn{2}{c|}{w/o GIC}                       & \multicolumn{2}{c|}{w/ GIC}                                                                            & \multicolumn{2}{c|}{w/o GIC}                       & \multicolumn{2}{c|}{w/ GIC}                                                                            & \multicolumn{2}{c|}{w/o GIC}                       & \multicolumn{2}{c}{w/ GIC}                                                                            \\ \midrule
        Datasets $\downarrow$       & {ACC} & {ASR} & ACC                                               & ASR                                               & {ACC} & {ASR} & ACC                                               & ASR                                               & {ACC} & {ASR} & ACC                                               & ASR                                               \\ \midrule
    Reserved   & 90.20                    & 1.22                    & 90.28(\textcolor{darkgreen}{0.08}) & 1.62(\textcolor{darkred}{0.40})    & 86.53                   & 0.23                    & 87.35(\textcolor{darkgreen}{0.82}) & 0.48(\textcolor{darkred}{0.24})    & 86.79                   & 0.68                    & 87.03(\textcolor{darkgreen}{0.24}) & 0.77(\textcolor{darkred}{0.09})    \\
    Brightness & 89.72                   & 1.58                    & 89.92(\textcolor{darkgreen}{0.20}) & 1.68(\textcolor{darkred}{0.10})    & 86.31                   & 0.53                    & 86.76(\textcolor{darkgreen}{0.45}) & 1.52(\textcolor{darkred}{0.99})    & 87.58                   & 1.73                    & 88.57(\textcolor{darkgreen}{0.99}) & 1.76(\textcolor{darkred}{0.02})    \\
    Synthetic  & 87.82                   & 0.83                    & 89.95(\textcolor{darkgreen}{2.13}) & 1.56(\textcolor{darkred}{0.72})    & 85.04                   & 0.29                    & 86.42(\textcolor{darkgreen}{1.38}) & 0.90(\textcolor{darkred}{0.61})    & 88.50                    & 0.42                    & 88.65(\textcolor{darkgreen}{0.15}) & 0.53(\textcolor{darkred}{0.11})    \\
    External   & 79.66                   & 0.80                     & 86.99(\textcolor{darkgreen}{7.33}) & 1.31(\textcolor{darkred}{0.51})    & 77.55                   & 0.29                    & 84.35(\textcolor{darkgreen}{6.80}) & 1.02(\textcolor{darkred}{0.73})    & 79.87                   & 0.78                    & 83.98(\textcolor{darkgreen}{4.11}) & 1.22(\textcolor{darkred}{0.44})    \\ \bottomrule
    \end{tabular}}

\end{table*}
In this section, we conduct a series of experiments to demonstrate the efficacy of GIC, testing it against multiple backdoor attacks and across a range of defense methods.

\textbf{Experiment settings.} In our experiments, we follow the attack and defense methodologies outlined in Section~\ref{sec::impact}. For the GIC method, we use a simple perturbation function \( g(x) = x + \epsilon \), where \( \epsilon \) is a sample-specific learnable perturbation. The perturbations are optimized using the objective defined in Equation~\ref{bce1}. To prevent the learning of unintended noise or malicious patterns with large norms, we constrain \( \epsilon \) with an $L_{\infty}$ norm less than 0.1. The perturbations are learned using the Adam optimizer with a learning rate of 0.1 over 100 steps, starting from zero initialization. This setup ensures that the perturbations remain within reasonable bounds, optimizing their alignment with the victim model’s expectations. The perturbation function and hyperparameters (e.g., learning rate and range of \( \epsilon \)) can be further fine-tuned to optimize GIC's performance, and we leave a more comprehensive study of them for future work. To maintain clarity, we present the most representative experiments in the main text, focusing on the key insights into GIC's performance across different configurations. The remaining experiments, which explore additional datasets, auxiliary data types, models, and attack-defense combinations, are deferred to the \app.

\begin{table}[ht]
\centering
    \caption{Experiments for FT-SAM with $\mathcal{D}_{ext}$ against various attacks. Both ACC and ASR for auxiliary without (w/o) applying GIC and with (w/) applying GIC are reported.}
    \label{gic_defense}
    \setlength{\tabcolsep}{6.00pt} 
    \renewcommand{\arraystretch}{1} 
    \scalebox{0.85}{
\begin{tabular}{c|cc|cc|cc}
\toprule
Attack $\rightarrow$                 & \multicolumn{2}{c|}{BadNets}                        & \multicolumn{2}{c|}{Blended}                         & \multicolumn{2}{c}{Input-Aware}                     \\ \midrule
                  GIC $\downarrow$    & {ACC} & {ASR} & ACC                       & {ASR} & ACC                       & {ASR} \\ \midrule
w/o GIC & 79.66                   & 0.80                    & \multicolumn{1}{r}{80.45} & 10.07                   & \multicolumn{1}{r}{85.13} & 1.58                    \\
w/ GIC    & 86.99                   & 1.31                    & \multicolumn{1}{r}{88.71} & 10.47                   & \multicolumn{1}{r}{90.54} & 1.36                    \\ \midrule \midrule
Attack $\rightarrow$                & \multicolumn{2}{c|}{SIG}                           & \multicolumn{2}{c|}{LF}                              & \multicolumn{2}{c}{WaNet}                           \\ \midrule
     GIC $\downarrow$             & {ACC} & {ASR} & ACC                       & {ASR} & ACC                       & {ASR} \\ \midrule
w/o GIC & 80.87                   & 0.63                    & \multicolumn{1}{r}{82.36} & 3.32                    & \multicolumn{1}{r}{83.64} & 1.60                     \\
w/ GIC    & 85.13                   & 1.33                    & \multicolumn{1}{r}{89.77} & 3.53                    & \multicolumn{1}{r}{89.72} & 1.79                  \\ \bottomrule
\end{tabular}}
\end{table}
\textbf{Effectiveness of GIC.} Table~\ref{gic_badnet} presents the results of applying GIC to six representative defense methods against the BadNets attack on PreAct-ResNet18. The findings demonstrate the significant impact of GIC in enhancing backdoor purification across various auxiliary datasets and defense methods. For all tested defenses (ANP, NPD, FT, FT-SAM, FST, and SAU), GIC consistently improves the model's ACC. Notably, external datasets benefit greatly, with improvements of up to +8.64\% in FT and +7.33\% in FT-SAM. This indicates that GIC effectively calibrates auxiliary datasets to align more closely with the victim model's training distribution, thus improving their utility in backdoor purification.

However, while GIC improves ACC, it also introduces variations in ASR. In some cases, ASR increases, while in others, it decreases, indicating a complex tradeoff between ACC and ASR. As discussed earlier, shifts in the feature space can help in backdoor purification, but overly large shifts can harm the model’s utility. GIC's alignment of the auxiliary dataset with the training distribution reduces the benefits of these shifts, potentially leading to higher ASR, which comes at the cost of ACC improvement.

\begin{table}[ht]
\centering
\caption{Experiments for FT-SAM with $\mathcal{D}_{syn}$ against various attacks with different training dataset.}
\label{gic_data}
\setlength{\tabcolsep}{4.00pt} 
    \renewcommand{\arraystretch}{1} 
    \scalebox{0.85}{
\begin{tabular}{c|c|cc|cc|cc}
\toprule
\multicolumn{2}{c|}{Attack $\rightarrow$}                      & \multicolumn{2}{c|}{BadNets}                        & \multicolumn{2}{c|}{Blended}                         & \multicolumn{2}{c}{Input-Aware}                     \\ \midrule
                       & GIC                    & \multicolumn{1}{l}{ACC} & \multicolumn{1}{l}{ASR} & ACC                       & \multicolumn{1}{l}{ASR} & ACC                       & \multicolumn{1}{l}{ASR} \\ \midrule
\multirow{2}{*}{GTSRB} & w\textbackslash{}o GIC & 90.08                   & 0.43                    & {88.41} & 0.76                    & {93.42} & 0.89                    \\
                       & w\textbackslash GIC    & 96.62                   & 0.35                    & {92.02} & 1.35                    & {96.04} & 0.41                    \\ \midrule \midrule
\multicolumn{2}{c|}{Attack $\rightarrow$}                      & \multicolumn{2}{c|}{BadNets}                        & \multicolumn{2}{c|}{Blended}                         & \multicolumn{2}{c}{Input-Aware}                     \\ \midrule
\multirow{2}{*}{Tiny}  & w\textbackslash{}o GIC & 15.39                   & 0.04                    & {18.56} & 0.16                    & {22.02} & 1.95                    \\
                       & w\textbackslash GIC    & 39.62                   & 0.09                    & {43.06} & 0.36                    & {45.02} & 1.57                    \\ \bottomrule 
\end{tabular}}
\end{table}

\textbf{Effectiveness over various attacks and datasets.} 
Having evaluated GIC’s performance across various defense methods, we next investigate its effectiveness across different attacks and datasets. We evaluate GIC with the powerful FT-SAM purification technique using the most challenging external dataset $\mathcal{D}_{ext}$ for CIFAR-10 and $\mathcal{D}_{syn}$ for other datasets against various attacks. As summarized in Table~\ref{gic_defense} and Table~\ref{gic_data}, GIC consistently improves defense effectiveness, yielding significant improvements in ACC with minimal variations in ASR.

\textbf{Overall}, these results highlight GIC's adaptability and effectiveness in improving auxiliary dataset utility for backdoor defense, particularly when working with datasets exhibiting distributional shifts. GIC offers a simple yet powerful method for bridging the gap between auxiliary and training data, leading to more reliable backdoor mitigation.
\begin{figure}[h]
    \centering
    \includegraphics[width=0.6\linewidth]{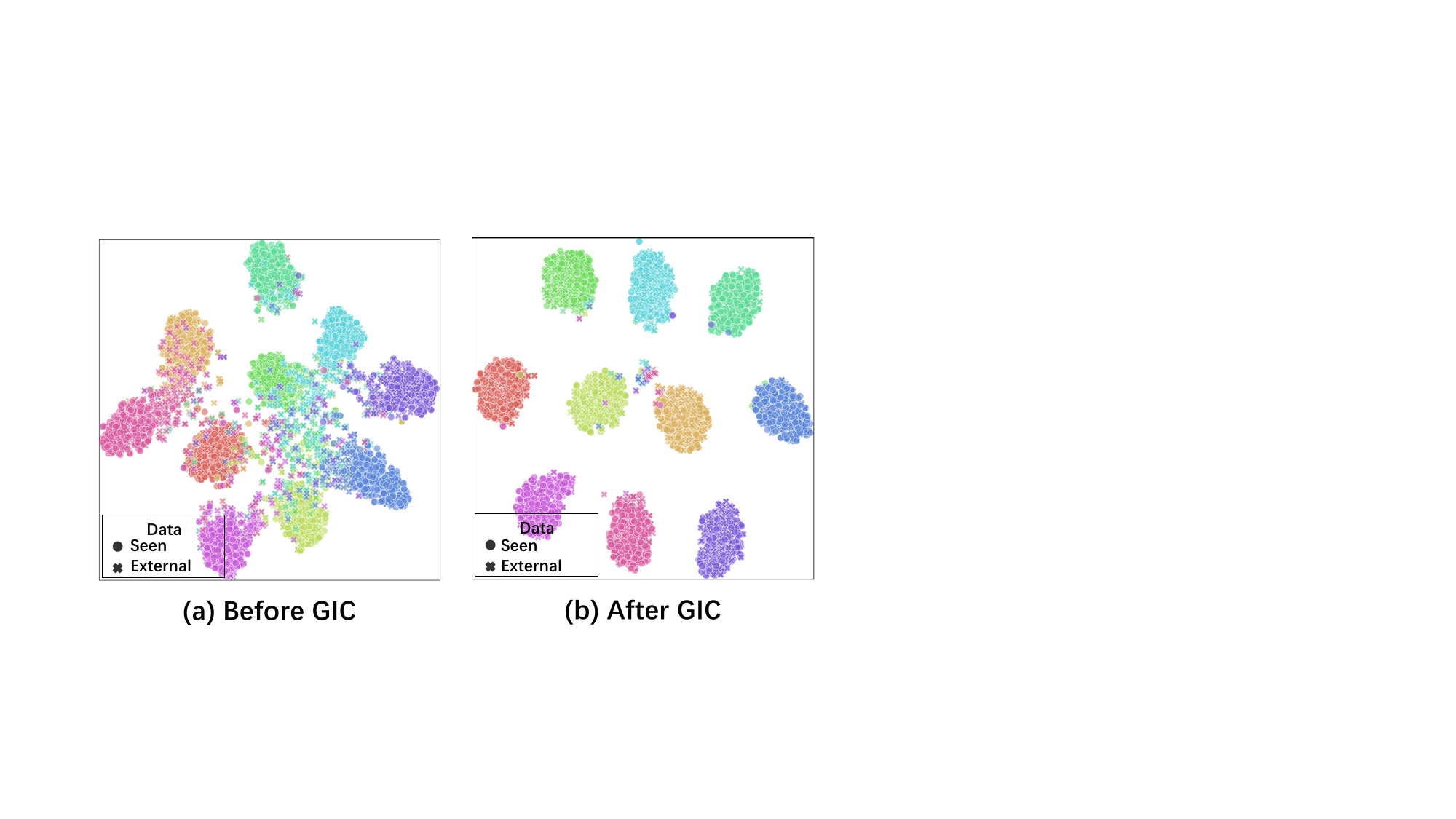}
    \caption{T-SNE visualization of feature representations for the external dataset, showing the transformation before and after applying Guided Input Calibration.}
    \label{tsnep2}
\end{figure}

\textbf{Understanding GIC.}  
To understand GIC, we visualize the features of external dataset before and after applying GIC. As shown in Figure~\ref{tsnep2}, the feature space analysis provides further insight into the role of GIC in aligning auxiliary data with the original training distribution. Without GIC, external data shows significant divergence from seen data in the feature space, highlighting the challenge of integrating out-of-distribution samples for backdoor purification. However, after applying GIC, the feature representations of the external data closely align with those of the seen data, demonstrating GIC’s effectiveness in transforming auxiliary samples to resemble the training data. This alignment confirms the theoretical foundation of GIC, as discussed in Theorem~\ref{thm}, where we showed that the transformation function \( g \) minimizes the discrepancy between auxiliary and training samples. 
The fact that GIC leads to nearly identical feature representations indicates that the optimization of \( g \) successfully adjusts the auxiliary data to match the high-confidence predictions of the victim model, ensuring better compatibility for backdoor purification. This feature space alignment not only enhances the purification process but also reduces the risk of reinforcing backdoor patterns by making the auxiliary data more consistent with the clean training distribution.

\section{Conclusion}
In this paper, we address a critical yet underexplored problem in backdoor purification, examining how different types of auxiliary datasets impact the effectiveness of state-of-the-art backdoor purification methods. Our findings underscore the pivotal role of auxiliary datasets in strengthening defenses against backdoor attacks, while also highlighting that not all datasets contribute equally to the purification process. Specifically, we found that in-distribution data preserves model utility but is less effective at neutralizing backdoors. In contrast, introducing entirely new datasets can cause the model to forget critical knowledge. 
To address these challenges, we propose Guided Input Calibration, an innovative approach designed to improve the efficiency of backdoor purification. GIC aligns the characteristics of auxiliary datasets with those of the original training set through transformations guided by the intrinsic properties of the compromised model. Extensive experiments demonstrate that GIC significantly improves the effectiveness of backdoor mitigation across various auxiliary datasets, making it a valuable tool for strengthening ML security in practical applications.

\textbf{Limitations and future work.} While our method offers promising results, we acknowledge several limitations and suggest avenues for future research. First, although GIC facilitates better alignment between auxiliary and original datasets, it still relies on some degree of similarity or the availability of relevant auxiliary data—a condition that may not always be feasible. Second, our method has been primarily tested on static ML models, leaving questions about its applicability to online learning environments or continuously updated models. For future research, we advocate exploring strategies that can function effectively without auxiliary datasets or with minimal data needs. Additionally, further investigation is needed into the role of auxiliary datasets for backdoor defense in other stages of ML systems, such as the pre-training and in-training stages.

\vspace{-0.1in}
\section*{Impact Statement}
Our primary objective is to develop a robust framework capable of effectively mitigating the complex threats associated with backdoor attacks, ensuring the integrity and dependability of applications based on ML. This work has significant positive implications for society, contributing to the protection of ML systems against malicious exploitation and promoting a safer digital environment.

\newpage
\bibliographystyle{plainnat}
\bibliography{references}

\end{document}